# Evolution of Galactic Nuclei with 10 $M_\odot$ Black Holes

*by*

Hyung Mok Lee

*Dept. of Earth Sciences, Pusan University, Pusan 609-735, Korea\**
*and*
*ITP, University of California, Santa Barbara, CA 93106, U. S. A.*

e-mail: hmlee@astrophys.es.pusan.ac.kr

\* *Present address*

## Abstract

A star with main sequence mass greater than $25 \sim 30\,M_\odot$ may collapse to a black hole of about $10\,M_\odot$ at the final stage of the evolution. About an order of 1% of stellar mass is likely to be in form of such black holes in galaxies. The presence of even such a small amount of 10 $M_\odot$ black holes in a spherical stellar system mainly composed of low mass main-sequence stars (i.e., $M \lesssim 1\,M_\odot$) significant effects on the dynamical evolution. We have examined the dynamics of two-component stellar systems composed of $0.7\,M_\odot$ main-sequence stars, representing the old population of stars whose main-sequence lifetimes are longer than the Hubble time, and a small fraction of $10\,M_\odot$ black holes. The dynamical friction leads to the segregation of black holes to the core and the core collapse takes place among the black holes in a time scale much shorter than that required for a single component cluster. Various physical processes occur as the density of the central cluster of black holes becomes higher: formation of binaries by three-body processes, heating by these binaries, capture between black holes by gravitational radiation and subsequent merger, tidal capture or disruption of stars by black holes. The ultimate evolution of the two-component stellar system depends on the role of three-body binaries formed among the black holes. In low velocity dispersion systems (i.e., $v \lesssim 100\,km/sec$ where $v$ is the three dimensional velocity dispersion), three-body binaries are eventually ejected from the stellar system after being hardened to hardness of a few hundred. This picture has to be altered as the velocity dispersion changes. For a system with $v \gtrsim 100\,km/sec$ binaries merge by gravitational radiation at some hardness instead of being ejected. The critical hardness, at which the collision time and the merger time become comparable, determines the efficiency of the binary as a heat source. The efficiency is found to be inversely proportional to the velocity dispersion. For the clusters without serious reduction in heating efficiency (i.e., velocity dispersion well below $500\,km/sec$), heating by three-body binaries have the effect of stopping the core-collapse. The cluster expands, but at a rate set by the half-mass relaxation time of the whole system which is very long. Thus one obtains nearly static two-component configuration: central cluster of black holes surrounded by low mass clusters. However, such a state would not last longer than Hubble time if $v \gtrsim 50\,km/sec$ because most of the black holes would experience binary formation and subsequent mergers. Thus a seed black hole can easily form in the central parts of galaxies with even moderate initial conditions (i.e., $v_c \gtrsim 100\,km/sec$).

## 1. Introduction

There are many ways to form massive black holes in the galactic nuclei (Rees 1984). Dynamical evolution of a stellar system mainly composed of neutron stars to a very high density and subsequent build up of individual objects through gravitational radiation loss has been pursued by a number of investigators (e.g., Quinlan & Shapiro 1989, and references therein). There are basically two constraints in order to form a central black hole through dynamical evolution: (1) Dynamical evolution time scale must be much shorter than the Hubble time, and (2) The collapse should continue until a central object forms. In order to satisfy these conditions with neutron star clusters, the cluster has to have $N \gtrsim 10^7$ and $v \gtrsim 500\,km/sec$ (Quinlan & Shapiro 1989). However, it is rather difficult to imagine how such a system can arise naturally. For example, when neutron stars are formed they are not the most massive components and mass segregation process actually acts against the formation of a very dense system of neutron stars.

It is also possible to form a very massive star, which turns into a black hole through dynamical instability, by successive mergers of normal stars (Quinlan & Shapiro 1990). However, one needs a rather fine tuned initial condition in order to avoid expansion before the formation of a massive star. On-the-spot star formation out of material ejected from stellar evolution would obviously enable continuous collapse but strong tidal field within the dense stellar systems can easily prevent from star formation activity within the dense parts.

Such difficulties can be alleviated by the presence of a small amount of black holes of several solar mass (i.e., much larger than individual mass of stars which comprises most of the mass). Recent study (Brown & Bethe 1994) indicates that there are two distinct types of black holes that are formed at the end of the stellar evolution: 1.5 $M_\odot$ black holes which result from the supernova explosion of 18–30 $M_\odot$ (main-sequence mass), and 10 $M_\odot$ black holes formed by the collapse of helium core of stars more massive than $25 \sim 30\,M_\odot$. The arguments based on the production of heavy elements compared to the production of helium through stellar evolution also give 25 $M_\odot$ for the stellar mass above which direct collapse to a black hole will occur (Maeder 1992, 1993). About $0.5 \sim 1\%$ of the stellar mass is estimated to be in the form of 10 $M_\odot$ black holes assuming the Salpeter initial mass function (Bethe & Brown 1994).

The dynamical effects of such black holes are discussed by Kulkarni, Hut & McMillan (1993) for the case of globular clusters. Because there could be only a small number of black holes (i.e, up to $\sim 100$ in rich clusters) in any cluster, the subsystem composed of black holes formed through mass segregation will not survive for a long time. The black holes will be ejected through the formation of binaries and interaction with others in several crossing times, which is very short. Thus one expects only a few, if any, black holes to remain in globular clusters at present time.

The situation in more massive stellar systems is quite different. A galactic nucleus is embedded within a much larger galaxy, but the extent within which a significant dynamical evolution is expected to be about a few parsec with enclosed mass being about $10^{7-9}\,M_\odot$. There are actually several galaxies which have distinct 'nuclei' within the flat 'core'. The typical nuclei density is around $10^{6-7}\,M_\odot\,pc^{-3}$ with enclosed mass being $10^{7-8}\,M_\odot$ (e.g., Lauer 1989). Therefore the number of black holes that reside within such a nucleus can easily exceed $10^4$ even with 1% of the mass fraction. It is clear that the role of 10 $M_\odot$ black holes would be very important in the dynamics of galactic nuclei if the subsystem of black holes can be formed well within the Hubble time. Furthermore, since the galactic nuclei are not isolated systems, continuous injection of black holes from increasingly larger radii is possible through the dynamical friction (Morris 1993).

The purpose of the present paper is to examine various aspects of the evolution of stellar systems that may be appropriate for the nuclei of galaxies, composed of two species: 0.7 $M_\odot$ main-sequence stars and 10 $M_\odot$ black holes that are formed by stellar evolution. We are especially interested in the possibility of the formation of a seed black hole through the dynamical evolution. Subsequent growth to a super massive black hole ($M \sim 10^{6 \sim 9} M_\odot$) in galactic nuclei is a separate question which is beyond the scope of the present paper.

A two-component configuration described above will arise in a relatively short time scale because of steep dependence of stellar life-time on mass. Unlike the neutron stars, 10 $M_\odot$ black holes would become the most massive component in $\sim 10^7$ years unless continuous star formation activity goes on.

We describe various aspects of the dynamical evolution of two-component clusters in §2. Some numerical results based on Fokker-Planck calculations are presented in §3. Possible scenarios for the evolution of various types of galactic nuclei are presented in §4 based on our results in §2 and §3. The next section concerns with the possible effects that might present in the realistic clusters. The summaries are given in the final section.

## 2. Dynamical Evolution of Two-Component Clusters

While two component representation is a great simplification for the actual stellar systems such as globular clusters or galactic nuclei, one can get a clear picture by studying simpler models. We expect that the major points will remain the same in more complex models with realistic mass spectrum. We assume that the model clusters are composed of 0.7 $M_\odot$ main-sequence stars, that represent the old stellar population of lifetime longer than the Hubble time, and 10 $M_\odot$ black holes, that are presumed to have formed by the collapse of massive first generation of stars in the galaxies. The mass of black holes formed by the collapse of the helium core should have a range of mass (between 5 and 20 $M_\odot$: Woosley & Weaver 1986). We adopt 10 $M_\odot$ as a representative value, but the presence of the range of mass will only make the stellar system more vulnerable to the rapid collapse. Changing the individual mass ratio will result in the changes in the evolutionary time scale but the qualitative results will not change much.

We will also limit ourselves to the Plummer model as an initial model for simplicity. Different initial configuration will evolve in somewhat different time scales, but the the general behavior of the physical quantities in the central parts will not be affected. The distribution of both main-sequence stars and the black holes are assumed to be the same in the initial model.

## 2.1 Mass Segregation and Core collapse

The core collapse of single mass component clusters has been studied in great detail by Cohn (1980). For an initial cluster of Plummer model, the core collapse takes place in 15.7 $t_{rh,0}$ where $t_{rh}$ is the half-mass relaxation time defined as

$$t_{rh} \equiv \frac{0.138 N^{1/2} r_h^{3/2}}{\langle m \rangle^{1/2} G^{1/2} ln(0.4N)}. \quad (1)$$

Here $N$ is the total number of stars, $r_h$ is the half-mass radius, and $\langle m \rangle$ is the mean mass of individual stars.

The core collapse will be greatly accelerated if there exists a massive component as we have argued earlier. The core collapse for multi-component systems has been studied by Inagaki (1985), but these were restricted to the cases with relatively small mass ratio ($m_2/m_1 \leq 5$). Most notable effect was the acceleration of the core collapse compared to the single component case. Such an acceleration is a combined result of mass segregation, which can be completed within $m_s/m_B t_{rel}$, and collapse of the high mass component whose relaxation time is much shorter than the relaxation time for the entire system.

We have shown the quantitative behavior of the time to complete core collapse in Figure 1a in units of initial half-mass relaxation time as a function of the fraction of massive black holes by mass. It clearly shows a broad minimum for the core collapse time indicating that large amount of acceleration of core collapse is possible for a wide range of black hole fraction. Even a small amount of black hole is enough to accelerate the core collapse by an order of magnitude compared to the single component case.

Since the relaxation time is different for the clusters with the same total mass and initial velocity dispersion but different mixture because of $<m>$ in eq [1]), it may be useful to measure the time to core collapse in absolute time. Thus we have normalized the time to core collapse in units of the initial half-mass dynamical time which is defined ae

$$t_{dh} \equiv \frac{r_h}{\sigma_c}, \quad (2)$$

where $\sigma_c$ is one dimensional velocity dispersion in the core.

The time to collapse in this unit is shown in Figure 1b, where we have assumed that the cluster has the total mass of $3 \times 10^7 \, M_\odot$, and initial one-dimensional velocity dispersion of 100 $km/sec$. These two parameters give the initial half mass dynamical time of $2.8\times10^4$ years. The core collapse time in this unit also has the broad minimum. This is a clear indication for the importance of dynamical friction in accelerating the core collapse. Note that the core collapse is faster for mixture of black holes and main-sequence stars than the case with black holes only, where the half-mass relaxation time is shortest. We also emphasize that the core collapse is rather limited to the black holes and the low mass stars undergo much small amount of dynamical evolution.

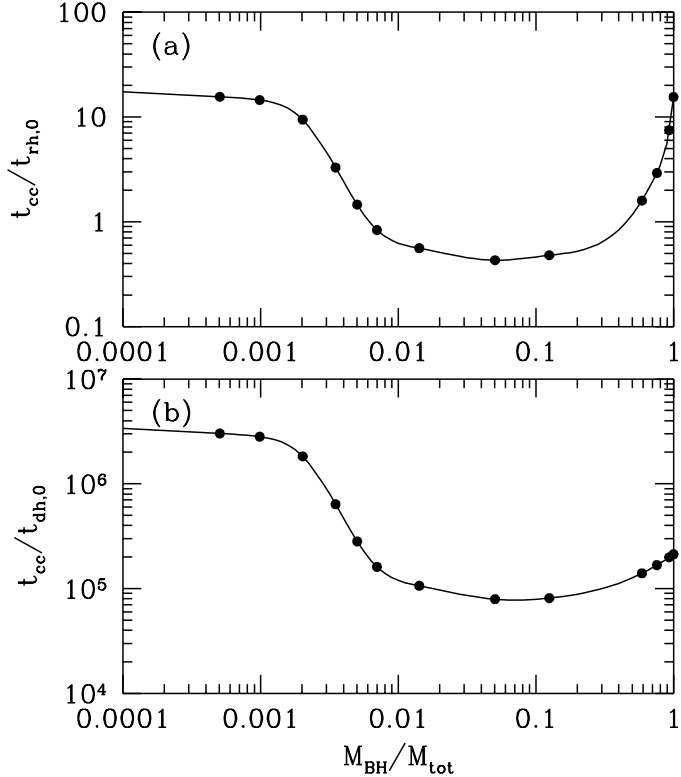

**Fig. 1** - The time to reach the complete core collapse without any heating mechanism measured by (a) initial half-mass relaxation as defined in eq. (1), and by (b) half-mass crossing time, as a function of mass fraction of black holes for the two component cluster composed of 0.7 $M_\odot$ (main sequence) and 10 $M_\odot$ (black hole) stars. The initial configuration is assumed to be a Plummer model.

Through the energy exchange between high and low mass components, approximate equipartition is established, although there is some deviation from the equipartition which drives the mass segregation instability (e.g., Spitzer 1987). After the mass segregation is almost complete (that is, when the half-mass radius of the black holes does not shrink further), the inner part can be regarded as a self-gravitating subsystem (mainly composed of black holes) which is in virial equilibrium, i.e.,

$$0.4\frac{GM_B}{r_{h,B}} \approx v_B^2, \qquad (3)$$

where $M_B$, $r_{h,B}$, and $v_B$ are total mass, half-mass radius, and velocity dispersion of the black holes, respectively. The surrounding system of main-sequence stars can also be characterized by the virial equilibrium

$$0.4\frac{GM}{r_h} \approx v^2, \qquad (4)$$

where $v$ is the *density weighted* velocity dispersion of stars. In the limit with $M_B/M \ll 1$, $v^2 \approx v_s^2$, and $\xi m_B v_B^2 \approx m_s v_s^2$, we obtain $r_{h,B} \approx \xi(M_B/M_{tot})(m_B/m_s)r_h$, where $\xi(\lesssim 1)$ is a measure of departure from the equipartition. On the other hand, if the fraction of heavy component is significant, $v^2 \approx v_B^2$. In that case, we obtain $r_{h,B} \approx r_h(M_B/M_{tot})$. Thus $r_{h,B}$ is much smaller than $r_h$ if $M_B/M_{tot}$ is much less than $m_B/m_s$. For our typical case of $M_B/M_{tot} = 0.01$ and $m_B/m_s \approx 14.3$, we have $r_{h,B} \approx 0.14\xi r_h$ with $\xi \approx 0.7$.

Since the surrounding cluster of low mass stars has a large core radius which changes very slowly, the half-mass radius of black holes become smaller than the core radius of normal stars. Thus the black holes become embedded in a *uniform* background of normal stars within $t \ll t_{rh,0}$.

### 2.2 Formation and Evolution of Three-body Binaries

The core collapse is a process that leads to an infinite density in a finite time. In practice, infinite density cannot be realized because of discrete nature of stars. If the density becomes sufficiently large, binaries can be formed through three-body processes. These binaries will subsequently interact with field stars (black holes in the present case) and release energy to the stellar system by hardening their orbits until they are ejected.

The rate of formation of three-body binaries can be written (Hut and Goodman 1992),

$$\frac{dn_{3B}}{dt} = 126 G^5 m^5 n^3 v^{-9}. \qquad (5)$$

We can immediately see that the three-body binaries among the black holes will be formed efficiently because of strong dependencies on the velocity dispersion, which is expected to be smaller for black holes within a stellar system of lower mass, and on the individual mass.

The binaries are formed with moderate hardness ($x \equiv \frac{3E_B}{mv^2} \approx 3$), and continuously harden. The final fate of binaries formed by three-body processes will be dependent on various combination of external (i.e., velocity dispersion, densities, etc.) and internal (i.e., mass of individual particles) parameters. The binaries eventually have to be ejected when typical recoil velocity exceeds the escape velocity from the stellar system. The expectation value for $y [\equiv \Delta E_B/(1/3mv^2)]$ is about $0.4x$, where $\Delta E_B$ is the typical change in binding energy of the binary per strong collision. About 1/3 of the binding energy change will be become translational energy of the binaries for the case of identical masses for both binary components and single stars. For an isolated cluster with central potential depth of $\chi \times 1/3mv^2$, one expects that the binaries will be ejected at $x \approx \frac{3}{0.4}\chi$. Typical globular clusters clusters reach $\chi \approx 14$ before the core begins to expand. This means that $x$ can become about 105. The total energy given to the cluster per binary is then $102 \times \frac{1}{3}mv^2 (= 102kT)$.

Such a picture would change in galactic nuclei with 10 $M_\odot$ black holes in two respects. If the depth of the central potential is measured by the typical kinetic energy of *black holes* whose mean kinetic energy is much smaller than the virial value (by about a factor of $\frac{1}{\xi}\frac{m_s}{m_B} \approx \frac{1}{14\xi}$), $\chi$ becomes very large. In the numerical calculations reported later, $\chi$ becomes as large as 60. This translates into the *enhancement* of the heating amount per binaries by about a factor of four. On the other hand, the orbit of hard binaries are expected change due to gravitational radiation.

The time scale to reach a complete merger for a an equal mass binary with elliptical orbit can be written as

$$\tau_{gr} = g(e) a^4 \frac{5c^5}{512 G^3 m^3} = g(e) \frac{405}{8192} \frac{Gmc^5}{x^4 v_c^8}, \qquad (6)$$

where $g(e)$ is sensitively varying function of $e$ (especially near 1) and is shown in Fig. 2 of Peters (1964). For the eccentricity distribution of $f(e) = 2e\,de$, the average value for $g(e)$ is about 0.22.

A necessary condition for a binary to become a heat source is that $\tau_{gr}$ should be longer than typical collision time scale ($\tau_{coll}$). In order to estimate $\tau_{coll}$, we consider the collisions with $r_p \lesssim a$, where $r_p$ is the distance of closest approach between a single and center-of-mass of the binary. Then the cross section can be written

$$\sigma_{coll} \approx \pi r_{p,max}^2 \left[1 + \frac{2G(m_b + m_s)}{v_{rel}^2 r_{p,max}}\right], \qquad (7)$$

where $m_b$ and $m_s$ are masses of binaries and singles, respectively, and $v_{rel}$ is the relative velocity at infinity. For binaries with $x \gg 1$, the gravitational focusing term will dominate. By assuming $r_{p,max} \approx a$ for strong encounters, we can estimate the collision time as

$$\tau_{coll} \approx (n_s <\sigma_{coll} v_{rel}>)^{-1}, \qquad (8)$$

where the square brackets represent the average taken over the velocity distribution. If we assume Maxwellian velocity distribution for both binaries and singles, the distribution of relative velocity is also Maxwellian with the rms velocity being

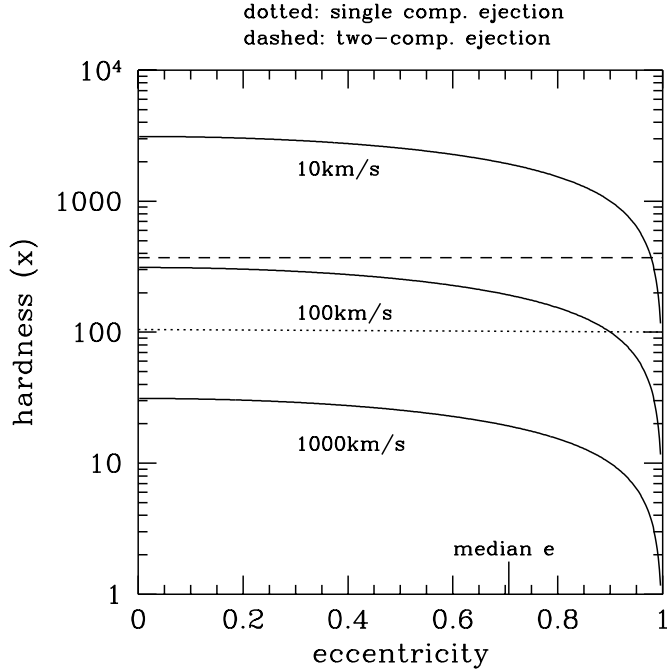

**Fig. 2** - The plot of the lines where the merger time scale is equal to the 'strong' collision time scale on eccentricity-hardness plane for different values of the velocity dispersion. For a given eccentricity, a three-body binary become mergers before experiencing any strong (i.e., energy exchanging) interaction with other black holes. Also indicated are the hardness at which any strong interaction leads to the ejection of the binary. The dotted line is the case for single component while the broken line is the case for two component (i.e., $0.7\,M_\odot$ normal stars and $10\,M_\odot$ black holes). The difference in critical hardness stems from the fact that the central potential depth is different in units of the velocity dispersion.

$$<v_{rel}^2>^{1/2} = \left(v_s^2 + v_b^2\right)^{1/2} \approx \left(\frac{3}{2}\right)^{1/2} v_s, \qquad (9)$$

where $v_s$ and $v_b$ are, respectively, velocity dispersions of singles and binaries, and we have assumed equipartition between these two components.

The three-body binaries will be able to stop the core collapse only when the number of stars in the core becomes of order of $10^1$ while it reaches $\sim 10^2$ during self-similar expansion (e.g., Goodman 1987). The number of stars in the core can be estimated from

$$N_c \approx \frac{2\pi}{3} r_c^3 n_c \approx 2^{-3/2} \left(\frac{3}{2\pi}\right)^{1/2} \frac{1}{(Gm)^{3/2}} \frac{v_c^3}{n_c^{1/2}}, \qquad (10)$$

where we have used the King's core radius for $r_c$

$$r_c \equiv \left(\frac{3v_c^2}{4\pi G m n_c}\right)^{1/2}. \tag{11}$$

Therefore we can express $n_c$ (density in the core) in terms of central velocity dispersion ($v_c$), and $N_c$,

$$n_c \approx \frac{1}{N_c^2} \frac{3}{16\pi} \frac{v_c^6}{(Gm)^3}. \tag{12}$$

Above formula gives $n_c \sim 10^{13} pc^{-3}$ at the time of reversal of the core collapse if we assume that $v_c \approx 100\,km/sec$. The central velocity dispersion is always dominated by that of black holes when three-body binaries become important. Note that the central velocity dispersion changes rather slowly during the core collapse, but the central density at rebounce is very sensitively dependent on $v_c$.

Now we evaluate the collision time scale. First, $<v_{rel}^{-1}> = (6/\pi)^{1/2} <v_{rel}^2>^{-1/2} \approx \sqrt{\frac{2}{\pi}} v_c^{-1}$ for Maxwellian distribution of $v_{rel}$. Thus the collision time becomes,

$$\tau_{coll} \approx \frac{v_c^3}{18 G^2 m^2 \pi^{1/2} n_c} x \approx \frac{8}{27} \pi^{1/2} N_c^2 \left(\frac{Gm}{v_c^3}\right) x, \tag{13}$$

where we have replaced $a$ by corresponding $x$ and central velocity dispersion is assumed to be dominated by the (single) black holes. By equating $\tau_{coll}$ and $\tau_{gr}$, we obtain critical $x$ beyond which the merger takes place before strong encounters,

$$x_{crit} = 0.623 \, [g(e)]^{1/5} \left(\frac{c}{v_c}\right) \left(\frac{1}{N_c}\right)^{2/5}, \tag{14}$$

for a given $e$. If we take an average value of $g(e)$ (=0.22) and assume $N_c \approx 100$, the $x_{crit} \approx 230 \left(\frac{100\,km/sec}{v_c}\right)$. This means that the efficiency of three-body binaries can be higher than the that for a single component, isolated system for $v \lesssim 230\,km/sec$.

In Figure 2, we have shown the critical hardness at which gravitational radiation merger time scale equals to the strong encounter time scale as a function of eccentricity for different values of $v_c$. Also shown as straight lines are critical hardness beyond which black holes will be ejected as a result of a strong encounter. The dotted line assumes that $\chi = 14$ (a value for a single component cluster) and the broken line assume that $\chi = 60$ (a value for $0.7\,M_\odot$ plus $10\,M_\odot$ black holes). It is clear that most of the three-body binaries formed in a cluster whose velocity dispersion is greater than $100\,km/sec$ will become mergers. The possibility of mergers increases for two-component clusters than a single component ones. *Thus in the environment of typical galactic nuclei, three-body binaries release large amounts of energy to the cluster and eventually become mergers at the end.* We emphasize here that $v_c$ refers to that of black holes. The stellar velocity dispersion is typically about three times larger than $v_c$ for our mixture.

Above estimates for the critical hardness are likely to be upper limits. First, the eccentricities can change by rather distant encounters. Therefore, as soon as the highly eccentric binaries are depleted, they will be recovered at a rate faster than the rate of strong encounters. If we assume that the encounters with $r_p \lesssim 5a$ can reestablish the thermal distribution of $f(e) = 2e\,de$, the diffusion time scale for eccentricity space will be shorter than the collision time scale that we have assumed in equation (14) by about a factor of five. If we use this time scale to determine critical $x$, it will be smaller by about 30%. Second, the strong interaction between a binary and a single often lead to very close encounters between any of three black holes involved in the collisions. During this phase, the instantaneous gravitational radiation can be significantly large. It is then possible to form a merger during a single resonant encounter. The reduction of efficiency due to this process is difficult to estimate because one has to follow the numerical integration of motion of particles. We only note here that similar phenomenon can occur in globular cluster dynamics where binaries composed of main-sequence stars can interact with other stars. Because of finite size of the stars, the direct collision frequently occur during complex encounters. Given these uncertainties, we will simply assume that the efficiency of binary heating can be smaller by 30% than the estimate based on equation with $\langle g(e) \rangle = 0.22$, i.e. we use $x_{crit} \approx 170 \left( \frac{100\,km/sec}{v_c} \right)$.

### 2.3 Capture by Gravitational Radiation

When two black holes become close, energy loss due to gravitational radiation can exceed the orbital kinetic energy and binaries can form. The cross section of such a capture can be written (e.g., Quinlan & Shapiro 1989)

$$\sigma_{gr,cap} \approx 2\pi \left( \frac{85\pi}{6\sqrt{2}} \right)^{2/7} \frac{G^2 m_1^{2/7} m_2^{2/7} (m_1 + m_2)^{10/7}}{c^{10/7} v_{rel}^{18/7}}, \qquad (15)$$

where $m_1$ and $m_2$ are masses of two approaching black holes and we assumed that the gravitational focusing term dominates the geometrical term. The gravitational radiation merger time scale is very short even though the typical hardness is only about 15 because the orbits are extremely eccentric [i.e., $e \approx 1 - 3.1 \times 10^{-3} \left( \frac{v}{10^3\,km/sec} \right)^{10/7}$; Quinlan & Shapiro (1989)]. Once a capture takes place, the orbit of binaries decay to a lower binding energy and circular orbits through subsequent gravitational radiation. The perturbations by neighboring black holes would not be significant under our physical conditions (Quinlan & Shapiro 1989; Lee 1993).

We now compare the rates of formation of three-body binaries and direct two-body captures. For equal mass black holes, we apply the capture cross section in equation (15) to obtain the capture rate per volume, i.e.,

$$\frac{dn_{cap}}{dt} = \frac{1}{2} n_B^2 < \sigma_{gr,cap} v_{rel} > \approx 22.8 \frac{Gm_B^2}{c^{10/7}} v_c^{-11/7}. \qquad (16)$$

The ratio between the three-body binary formation and the gravitational radiation capture becomes

$$\frac{\dot{n}_{3b}}{\dot{n}_{cap}} \approx 0.66 \frac{1}{N_c^2} \left(\frac{v_c}{c}\right)^{10/7}. \tag{17}$$

Above equation has the same dependency on velocity and number of stars as that derived by Lee (1993) [eq. (11)], but has a significantly different constant (i.e., 18 versus 0.66). In our expression, we use the number of black holes in the core, while Lee (1993) used the total number. Our expression is based on the local conditions (i.e., central parts) while Lee's expression is based on 'global' conditions. Since the physical condition changes drastically over the location, we prefer to express in local quantities.

Equation (17) ensures that the three-body binary formation is more rapid than gravitational capture during the core-collapse and post-collapse phase as long as $v_c \lesssim 350 \, km/sec$ (with the assumption of $N_c \approx 100$). As emphasized earlier, the velocity dispersion is for the black holes and the stellar velocity dispersion is much higher. This means that one cannot ignore three-body binaries unless the stellar velocity dispersion is very large (i.e., $\gtrsim 1000 \, km/sec$).

### 2.4 Tidal Interactions Between Normal Stars and Black Holes

Tidal interaction between a normal star and a black hole can lead to the capture or instantaneous disruption of the normal star. The disruption will take place if the dimensionless parameter $\eta \lesssim 1$ (e.g., Kochanek 1993, but see Kosovichev & Novikov 1992 for slightly larger value for critical $\eta$), where

$$\eta \equiv \left(\frac{m_*}{M_B}\right)^{1/2} \left(\frac{r_p}{R_*}\right)^{3/2}. \tag{18}$$

This can be translated to the critical pericentral distance of $r_{dis} \approx R_* \left(\frac{M_B}{m_*}\right)^{1/3}$ ($\approx 2.4 R_*$ for encounters between $10 \, M_\odot$ black hole and a $0.7 \, M_\odot$ main-sequence star) for the immediate tidal disruption. More distant encounters will lead to the tidal capture. The critical pericentral distance for tidal capture is approximated to be $r_{tc} \approx 4 R_* \left(\frac{v}{100 \, km/sec}\right)^{-0.2}$ for a pair of $0.7 \, M_\odot$ main-sequence star and a $10 \, M_\odot$ point mass using the tabulated values of overlap integrals by Lee & Ostriker (1986), where $v$ is the typical relative velocity between a black hole and a main-sequence star. Because of the energy constraint, there wouldn't be any tidal capture if $v_{rel} \gtrsim 600 \, km/sec$. Therefore only a small fraction of strong tidal interactions will lead to the capture of the main-sequence star.

Even the captured stars will become disrupted as a result of repeated tidal interactions as the star continuously returns to the pericenter (see, however, Mardling 1994ab for other possibilities). It may be safe to assume that most of the strong tidal interactions will lead to the disruption of the main-sequence star.

The evolution of the disrupted material bears some importance both dynamically and observationally. The amount of the mass that will eventually be accreted by the black hole will determine the possible X-ray luminosity and speed of the growth of the black hole mass. For the dynamical point of view, the capture of normal star acts as an energy sink because the rapidly moving normal star becomes a part of slowly moving black hole. On the other hand, if some fraction of remnant gas is ejected from the black hole, and eventually from the central potential of the cluster, it will induce indirect heating. The heating per unit mass is $\phi_0$ while cooling per unit mass is $v_{rel}^2 (\approx v_*^2)$, where $\phi_0$ is the central potential, $v_{rel}$ is the typical relative velocity between black holes and stars, and $v_*$ is the stellar velocity dispersion. Therefore, the amount of heating per unit mass is always greater than the amount of cooling per unit mass (i.e., $\phi_0 \gg v_*^2$). It is thus important to know what fraction of the stellar material will be ejected. We now make a simple estimate for the remnant fraction.

Suppose that the captured star will be destroyed at a pericentral passage of $r_p$. At the time of the disruption, one can express the velocity of each particle as a superposition of the 'orbital velocity' and the random velocity, i.e.,

$$v = v_{orb} + \Delta v, \tag{19}$$

The binding energy per unit mass can be written as

$$E_b = \frac{1}{2} \left( v_{orb} + \Delta v \right)^2 - \frac{Gm_B}{r} \tag{20}$$

The root-mean-square velocity of the random component can be estimated from the requirement of

$$\frac{1}{2} (\Delta v)^2 \approx \frac{Gm_*}{R_*}, \tag{21}$$

which gives $(\Delta v)_{rms} \approx 440 \, km/sec$. If the star is disrupted in the initial orbit which is hyperbolic, the total binding positive. On the other hand, if the star is disrupted during the circularization process, the binding energy just before the disruption is negative and would be order of the stellar binding energy itself. The fraction of particles with positive energy (and thus would eventually be escaping) will be dependent on the average binding energy at the time of the disruption. The average binding energy is related with the orbital velocity through

$$< E_b > = \frac{1}{2} v_{orb}^2 - \frac{GM_B}{r_p} = -\gamma \frac{Gm_*}{R_*}, \tag{22}$$

where we have characterized the orbital binding energy by a dimensionless number $\gamma$. The collision is highly hyperbolic if $\gamma < -1$ while elliptical if $\gamma > 0$.

The condition for a specific binding energy of a particle being positive can be written

$$(\Delta \tilde{v})^2 + (\Delta \tilde{v}) \tilde{v}_{orb} \cos\theta + \frac{\gamma}{2} > 0, \qquad (23)$$

where we have used the tilde in order to express velocities normalized by the by the escape velocity from the stellar surface ($v_{esc} \equiv \sqrt{2Gm_*/R_*}$), and $\theta$ is the angle between the orbital motion and the random motion. The orbital velocity is usually greater than $v_*$ for the cases with $M_B/m_* \gg 1$, and the particles with $\cos\theta > 0$ are likely to have $E_b > 0$ for $|\gamma| \ll 1$. Therefore if the star is disrupted in the initial encounter with $\frac{1}{2}v^2 \ll v_{esc}^2$ (parabolic encounter), about 50% of the material will be unbound immediately after the disruption. If the relative velocity at infinite was comparable or greater than $v_*$, almost all the material would have positive binding energy and most of the disrupted material will be unbound. Obviously no tidal capture is possible in this case and most of the disrupted material will escape from the black hole.

If we ignore small fraction of capture, parabolic disruptions will lead to the 50% ejection and 50% bound to the black hole. The net effect will be heating at a rate of $0.5\dot{m}(\phi_0 - 0.5v_*^2)$. If the encounter is hyperbolic, the heating rate will be $\sim \dot{m}\phi_0$. Either case, the heating efficiency is limited by the central potential. The heating rate per unit mass for three-body binaries is determined by the central velocity dispersion. But if $v_c \lesssim 400\,km/sec$, the heating efficiency of three-body binaries exceeds that for the ejection of stellar material. Unless the rate of tidal disruption greatly exceeds that of three-body binary formation, the indirect heating will not be important. However the potential depth will be sufficiently deep to prevent the complete ejection of the disrupted material if $v_c \gtrsim 400\,km/sec$. Therefore the tidal disruption will not play dynamically important role in any circumstances. We only note that the amount of mass to be bound to the black hole will influence the amount of radiation and the growth of the mass of black holes.

### 3. Numerical Results

We have seen that there are several dynamical processes to be considered in order to determine consequences of dynamical evolution of a two-component cluster composed of low mass main-sequence stars and relatively massive black holes. We now describe the numerical computations using Fokker-Planck method.

We have included the following processes explicitly in the Fokker-Planck computations reported below: formation, heating and subsequent merger of three-body binaries, and two-body gravitational radiation capture. We have computed the distribution functions of binaries by gravitational radiation capture using using the same method described by Statler,

Ostriker & Cohn (1987). We have assumed that the three-body binaries generate the energy as soon as they are formed and instantaneously become mergers. The delay between the formation, hardening and subsequent merger should not be important because the lifetime of these binaries are considered to be much shorter than typical dynamical age. In order to account the heating effect from both 10 $M_\odot$ and 20 $M_\odot$ black holes, we have adopted the heating formula given by Lee, Fahlman, & Richer (1991). Unlike the previous studies involving three-body binaries, we have included the distribution function of these binaries because they become mergers instead of being ejected. As we will see in actual numerical results, the growth of the black hole mass will be dominated by the merger of three-body binaries. There are several ways to include the three-body binaries. In the present calculations we have simply assumed that the distribution function of the three-body binary merger products conforms that of two-body gravitational merger products. Since three body binaries formation rate has different dependence on velocity and density, the actual distribution would be different from that of gravitational merger products. However, both types are strongly concentrated toward the core. Furthermore, dynamical relaxation would make these two types indistinguishable within a few central relaxation time. When the central parts become very dense, the relaxation time scale is much shorter than the time scale for the formation of three-body binaries. Therefore, the difference between the distributions of three-body binary merger products and the direct two-body merger products will not be significant. Thus we chose the same distribution of these merger products for the sake of simplicity (and also to save the computing time).

We have adjusted the efficiency of the three-body binary heating with the central velocity dispersion as discussed in §2.2. We have assumed two different values for $M_{BH}/M_{tot} = 0.01$ and 0.005. We have fixed the total mass at $3 \times 10^7 M_\odot$, but varied the initial velocity dispersion from 100 $km/sec$ to 400 $km/sec$. Because the efficiency of the three-body binaries become too small for $v_c \gtrsim 500\, km/sec$, such clusters may collapse without any resistance. We have summarized the model parameters in Table 1. The initial density and velocity distributions are assumed to be Plummer model in all cases. The important model parameters here are the velocity dispersion. The models A1 and A2 may represent those with very shallow potential, while models C1 and C2 may represent those of bright galaxies. The intermediate models B1 and B2 may be appropriate for normal galaxies such as the Milky Way, M31 or M32. We have indicated the time of core collapse in the final column to give an idea of the earliest time for the formation of a seed black hole.

The evolution of the Model B1 (i.e., $v_{c,init} = 200\, km/sec$, $M_B/M_{tot} = 0.01$) is displayed in Figure 3. We have shown the (a) central density, (b) velocity dispersion, (c) half-mass and core radii, (d) number of stars (black holes) in the core, (e) accumulated masses that are involved in the three physical processes shown in, and merger time scale (see below for the definition).

**Table 1.** Model Parameters

| Model | $M_{BH}/M_{tot}$ | $v_{c,0}$ ($km/sec$) | $\rho_{c,0}$ $M_\odot pc^{-3}$ | $t_{rh,0}$ (years) | time unit (years) |
|---|---|---|---|---|---|
| A1 | 0.01 | 100 | $2.63 \times 10^4$ | $2.4 \times 10^{10}$ | $7.3 \times 10^8$ |
| A2 | 0.005 | 100 | $2.63 \times 10^4$ | $2.4 \times 10^{10}$ | $7.3 \times 10^8$ |
| B1 | 0.01 | 200 | $1.68 \times 10^6$ | $3.0 \times 10^9$ | $9.1 \times 10^7$ |
| B2 | 0.005 | 200 | $1.68 \times 10^6$ | $3.0 \times 10^9$ | $9.1 \times 10^7$ |
| C1 | 0.01 | 400 | $1.08 \times 10^8$ | $3.7 \times 10^8$ | $1.1 \times 10^7$ |
| C2 | 0.005 | 400 | $1.08 \times 10^8$ | $3.7 \times 10^8$ | $1.1 \times 10^7$ |

Figure 3(a) shows that the core is quickly dominated by black holes. After the collapse, the central parts undergo high amplitude oscillation which is known as 'gravothermal oscillation'. More detailed behavior of the central densities of various components of the models shown in Figure 3 is illustrated in Figure 4. The core density during peak collapse is dominated by 20 $M_\odot$ black holes while the 10 $M_\odot$ component is more important during the 'low density' period. Also evident from this figure is that the relative importance of the 20 $M_\odot$ component grows with time due to the fact that the number of this component increases through the merger of 10 $M_\odot$ black holes.

The variation in central density is about four orders of magnitude, but the central velocity dispersion changes only by a small amount as shown in Figure 3(b). This is mainly because of extreme insentiveness of the velocity dispersion on the central density during the homologous core-collapse (e.g., Cohn 1980).

The half-mass radius of the black hole component remains nearly static (Fig. 3(c)). This is a different behavior from the single component cluster whose half-mass radius expands in a half-mass relaxation time. The black hole component should also expand but in a time scale set by the half-mass relaxation time of the entire cluster which is much longer than the half-mass relaxation time of the black hole. Thus in some extent, the energy generated by black hole cluster is absorbed by the entire cluster and the whole cluster remains nearly 'static'.

The number of 'stars' (initially mostly normal stars but later mostly black holes) in the core also oscillate by about two orders of magnitudes (Fig. 3(d)). Obviously this could make the efficiency of three-body binaries change rapidly with time, but we have used rather small values for $N_c (\approx 100)$ in estimating binary efficiencies because the three-body binaries would not be any of importance in low density (and thus high $N_c$) regime even if we use the high efficiency because the three-body binary formation rate is very small during that stage.

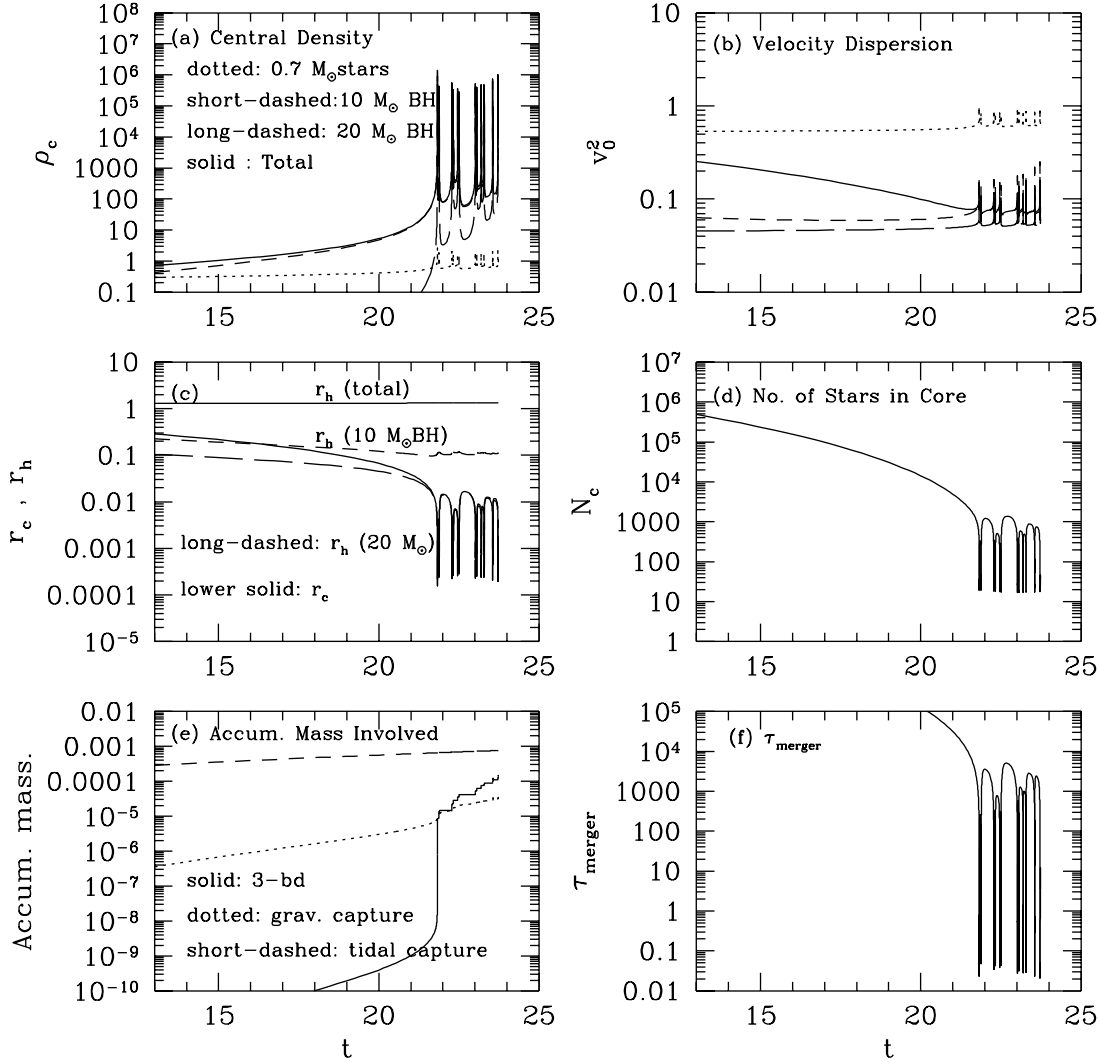

Fig. 3 - The evolution of (a) central density, (b) central velocity dispersion, (c) core and half-mass radii, (d) number of stars in the core, (e) accumulated masses involved in the tidal interaction, gravitational radiation merger, and three-body binary formation, and (f) runaway merger time scale (see eq. [24]) for model B1 (i.e., $M = 3 \times 10^7\,M_\odot$, $v_c(t=0) = 200\,km/sec$, $M_{BH}/M_{tot} = 0.01$. The following units are used: time in $9.1 \times 10^7$ years, density in $1.68 \times 10^6\,M_\odot/pc^3$, velocity dispersion in 283 $km/sec$, number of stars in actual numbers, accumulated mass in total mass (i.e., $3 \times 10^7\,M_\odot$).

The accumulated mass in Figure 3(e) (and later figures as well) needs some explanation. We have counted all the masses of 20 $M_\odot$ black holes formed by gravitational radiation capture and three-body binaries. However, we only counted the mass of main-sequence stars for the 'tidal capture' (which includes tidal disruptions) category. First to note here is that the amount of main-sequence stellar mass that has undergone the tidal interaction with the black hole exceeds the the other masses. That means that the black hole mass can grow most rapidly by accreting the main-sequence stars in an average sense. However since the main-sequence stars are almost uniformly distributed, the chance for a given black hole to tidally interact is nearly the same for all black holes. Therefore, by accreting the main-sequence stars, individual mass of the black holes would grow nearly uniformly regardless of the location of the black holes within the stellar system. Throughout our calculations, the total mass accreted by the black hole is about 10% of the entire black hole mass. This means that the each black hole would have grown by (at most) 10% in mass. This would not have any significant dynamical effect, and we have ignored the growth of the individual mass of the black holes by this process. However, this would be important if we consider the radiation from the stellar system because the mass accretion will result in radiation (presumably in UV or X-rays).

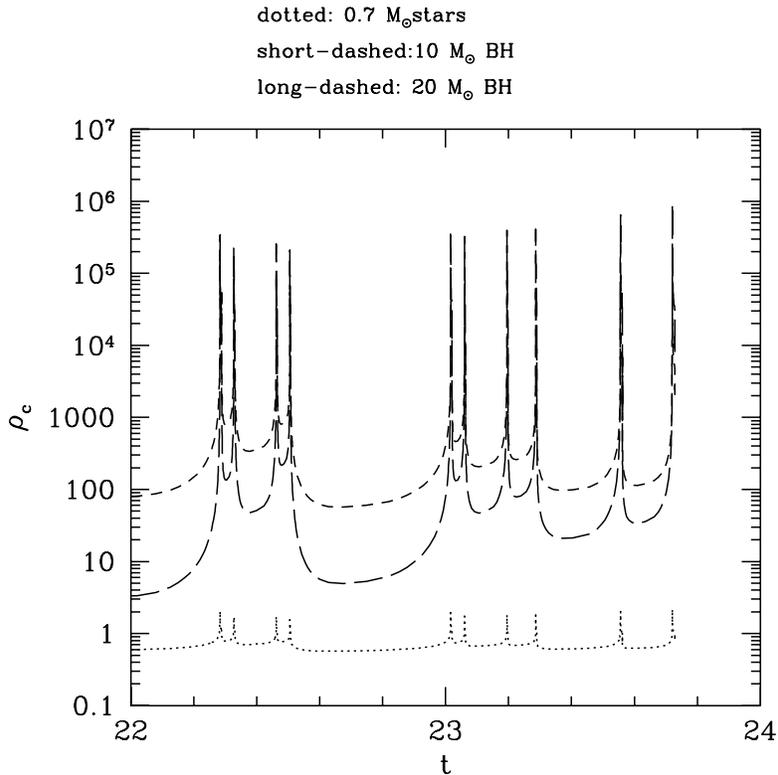

Fig. 4 The blown-up view of the evolution of the central density during the core oscillation for the model B1.

On the other hand three-body binary formation and subsequent merger and gravitational radiation capture are highly localized in the core. The accumulated mass participated in these processes are smaller than those in tidal capture, but the dynamical effects are larger. It is apparent that the number of merger products through the formation of three-body binaries exceeds that through direct two-body capture during the first core collapse. The subsequent oscillations tend to widen the gap between the products of these two processes.

We allowed only the formation of 20 $M_\odot$ black holes in the present numerical computations in order to maintain simplicity. More massive black holes would form through successive mergers (e.g., Quinlan & Shapiro 1989). Initially the mergers would generate 'ordered' mass function, but one single massive object forms later on (Lee 1993). Such an object forms in a 'merger time scale' defined as

$$\tau_{merger}^{-1} = \frac{n_c}{(dn_{3b}/dt + dn_{cap}/dt)}. \tag{24}$$

The formation for the single massive object is a result of stochastic process which cannot be modeled by Fokker-Planck method. We simply note that the necessary condition for the formation of such an object is that the age should be longer than the above merger time scale.

Since the central parameters are oscillating in time with rather large amplitude, the merger time scale defined above oscillates rapidly (see Fig. 3f). The amplitude of the oscillation is about four orders of magnitudes. Here, we make a simple estimate for 'average merger time scale' based on the average characteristics of the central parameters.

If we consider only three-body binary merger (which is dominant source for merger) the merger time scale becomes

$$\tau_{merger} = 0.45 \left(\frac{1}{Gm}\right) v_c^3 \frac{1}{N_c^4} \approx 4.2 \times 10^6 \left(\frac{N_c}{100}\right)^4 \left(\frac{m_B}{10\,M_\odot}\right) \left(\frac{100\,km/sec}{v_c}\right)^3 \; years, \tag{25}$$

where we have used the relationship between $n_c$ and $N_c$ from equation (12).

Obviously $N_c$ varies from 10 to 1000 for our actual runs. If we assume that $<N_c> \approx 100$, the merger time scale is much shorter than Hubble time. Again note that the velocity dispersion is for black holes, and smaller than the 'stellar' velocity dispersion by about a factor of 3. In all cases with $v_i \gtrsim 100\,km/sec$, the merger time scale is significantly shorter than the Hubble time, unless merger process is heavily weighted toward the high $N_c$. If we measure the time average value of $N_c$, it is on the larger side (i.e., close to 1000). However, since the merger rate is very steep function of $N_c$, low $N_c$ (i.e., high density phase) has higher weight. The choice of $\langle N_c \rangle \approx 100$ seems to be reasonable for estimating the merger time scale. We simply note that the merger time scale is very sensitive to the appropriate value of $N_c$, and there could be easily an order of magnitude uncertainty. Even with such a generous estimate for uncertainty, the case shown in Figure 3f will be able to sustain two-component phase only for less than $10^9$ years.

The evolution of model B2 (same as model B1 except the mass fraction of black holes is 0.5 %) is shown in Figure 5. The general evolution time scale is prolonged by about a factor of two. Nonetheless, it would be interesting to see the behavior of the system. After the core collapse, this model follows a qualitatively different path from model B1. The number of oscillations it experiences is small, and collapses indefinitely by converting most of the 10 $M_\odot$ black holes into 20 $M_\odot$ black holes. This is because it has roughly similar physical conditions (and thus merger rates) in the core as the model A1, but has smaller amount of black holes available. Larger fraction of black holes will be converted into 20 $M_\odot$ objects.

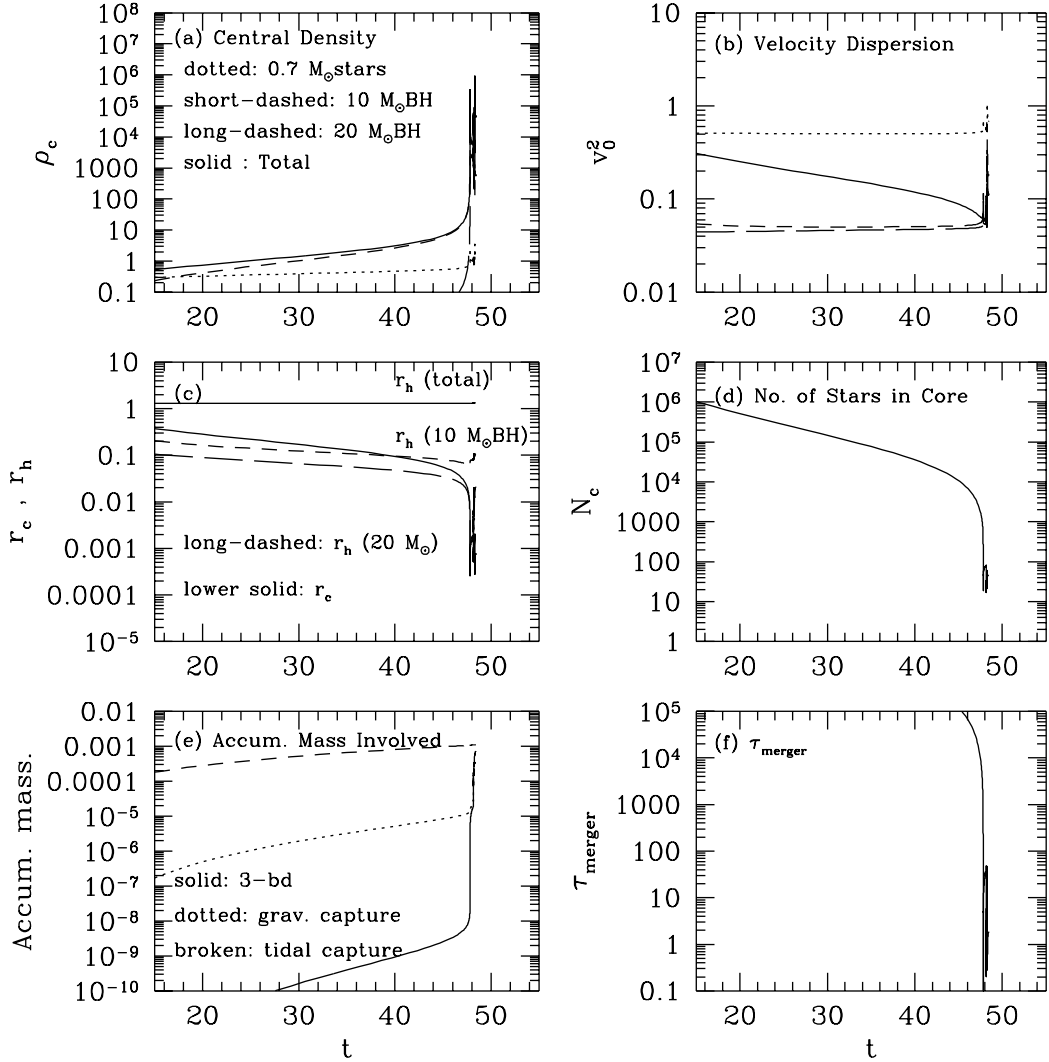

**Fig. 5** - The same as Fig. 3 for model B2 (i.e., same model parameters with B1 except that $M_{BH}/M_{tot} = 0.005$). All units are the same as in Fig. 3.

The models with lower velocity dispersions follow similar paths to those shown in Figures 3 and 5. The case with $v_i = 100\,km/sec$ (models A1 and A2) have similar behavior with models B1 and B2 except that the overall time scale is much longer. A seed black hole will form after experiencing some oscillations.

As the velocity dispersion becomes higher, larger fraction of black holes are converted into 20 $M_\odot$ object. The cases with $v_i = 400\,km/sec$ (models C1 and C2) experience core bounce (at very high central density), but the fraction of black holes experienced the merger becomes 20 % of the total black holes at the time of rebounce. This means that the oscillation may not develop, but a seed black hole can be formed during the first collapse.

## 4. Discussions

As we have seen in previous section, the presence of small fraction of 10 $M_\odot$ black holes can change the dynamical evolution of galactic nuclei toward the direction that the formation of a seed black holes can be accomplished with moderate initial conditions. This is because of the facts that the mass segregation takes place in a time scale shorter than general relaxation time and that formation of three-body binaries, which eventually become mergers, is efficient.

The heating by three-body binaries is able to stop the core collapse and gives rise to the two-component phase where expansion of the central cluster of black holes is supported by the energy absorption by the surrounding stellar system as long as $v_i \lesssim 500\,km/sec$. The central parts oscillate gravothermally. During this two-component phase rapid build up of black hole mass is possible in the core. The time scale required for the runaway merger is very short for $v_i \gtrsim 100\,km/sec$ once the central parts reached the core collapse.

The galactic nuclei are not isolated systems but part of the extended galaxy. Our assumption of isolated systems as models for galactic nuclei is clearly great simplification. We now discuss how we can apply our results to real galaxies.

Since the core collapse is limited to the massive component and it proceeds after the dynamical friction took place for a sizable amount of black holes, it would be instructive to derive what conditions are necessary in order to form a central subsystem of black holes in time scale shorter than Hubble time.

By applying the dynamical friction formula given by Binney & Tremaine (1987; eq. [7-24]) to a nearly uniform sphere composed of normal stars of mass $m_*$ and $m_B = 10\,M_\odot$ black holes, we obtain the 'friction time' within which a particle moving on a circular orbit at radius $r$ to spiral into the central parts,

$$t_{fric} = 0.53 \frac{(G\rho_*)^{1/2}}{G m_B ln\,\Lambda} r^3, \qquad (26)$$

where $\rho_c$ is the mass density of back ground stars (with mass $m_* \ll m_B$). By turning this around, we estimate the mass of the black holes to settle in the central parts by dynamical friction during the time $t$ by integrating the mass within $r$ where $t_{fric} = t$,

$$M_B = \frac{4}{3} f_B \pi r^3 \rho_* = 14.8\,(G\rho_*)^{1/2}\,ln\,\Lambda\,t m_B$$
$$\approx 3 \times 10^4\,M_\odot \left(\frac{f_B}{0.01}\right) \left(\frac{\rho_c}{10^5\,M_\odot\,pc^{-3}}\right)^{1/2} \left(\frac{ln\,\Lambda}{20}\right) \left(\frac{m_B}{10\,M_\odot}\right) \left(\frac{t}{10^{10} yrs}\right), \qquad (27)$$

where $f_B$ is the fractional mass of the black holes within the uniform sphere. It is not clear how much mass of the black holes are necessary in order to form a seed black hole, but $N_B$ (number of black holes in the central parts) probably should exceed a few thousand. Thus by requiring $M_B \gtrsim 3 \times 10^4 \, M_\odot$ within Hubble time the initial mass density of the uniform sphere has to be higher than $10^5 \, M_\odot \, pc^{-3}$. Most of the observed dense nuclei have $\rho_c \gtrsim 10^6 \, M_\odot \, pc^{-3}$ (Lauer 1989) and clearly satisfy this condition.

Similar analysis can be done for other extreme example of stellar density distribution of singular isothermal sphere (SIS). Now the friction time scale becomes

$$t_{fric} = 1.17 \frac{v_{rot} \ln \Lambda}{G m_B} r^2, \tag{28}$$

where $v_{rot}$ is the (constant) rotational velocity of the SIS. The mass of the black holes that would undergo dynamical friction within time $t$ now becomes

$$\begin{aligned} M_B &= 0.93 \frac{v_{rot}^{3/2}}{G^{1/2}} \frac{m_B^{1/2}}{(\ln \Lambda)^{1/2}} t^{1/2} \\ &= 3.9 \times 10^4 \, M_\odot \left(\frac{f_B}{0.01}\right) \left(\frac{v_{rot}}{250 \, km/sec}\right)^{3/2} \left(\frac{m_B}{10 \, M_\odot}\right)^{1/2} \left(\frac{20}{\ln \Lambda}\right)^{1/2} \left(\frac{t}{10^{10} yrs}\right)^{1/2}. \end{aligned} \tag{29}$$

Therefore, we now require $v_{rot} \gtrsim 250 \, km/sec$ in order to form a subsystem of the black holes of mass $\gtrsim 4 \times 10^4 \, M_\odot$ within an SIS.

The amount of black holes to undergo dynamical friction is proportional to $t$ for uniform sphere and to $t^{1/2}$ for SIS. If the actual nuclei are close to SIS outside the flat core, $M_B$ would grow proportional $t$ initially and proportional to $t^{1/2}$ after all the black holes in the flat core have undergone dynamical friction.

As we have seen from Figure 2, the nuclei with velocity dispersion smaller than $\sim 50 \, km/sec$ cannot hold the three-body binaries, but they rather are ejected at some hardness. Since the three-body binary formation is very efficient in low velocity dispersion environment, this process is likely to be efficient in removing black holes in stellar systems with small escape velocity, as discussed by Kulkarni, Hut & McMillan (1993) in the context of globular clusters. This may be the case for the central nucleus of M33, which has very little evidence for a massive black hole (Kormendy 1993).

The gravitational radiation during the final phase of merger between non-equal mass black holes contains non-zero linear momentum. This means that a merged black hole could have a recoil velocity. The maximum possible recoil velocity is estimated to be a few hundred km/s, but Redmount & Rees (1989) suggested a much larger recoil velocity could be possible if the black holes are rotating. Since there are many uncertainties on this subject, we neglected the recoil velocity in our numerical calculations reported earlier. However, it is clear that the growth of a black hole mass would be more difficult in clusters with small escape velocity (i.e., $v_{esc} \lesssim 200\,km/sec$).

As discussed in §2.1, subcluster of black holes would reside within a flat core of normal stars. Such a segregated state will last until a single massive black hole forms. Since the half-mass radius of the black hole subcluster is of order of

$$r_{h,B} \approx 0.8\,pc \times \left(\frac{M_B}{10^6\,M_\odot}\right)\left(\frac{100\,km/sec}{\sigma_*}\right)^2, \qquad (30)$$

the black hole subcluster will not be resolved in most galaxies. Here $\sigma_*$ is one dimensional velocity dispersion and we have assumed equipartition between black holes and normal stars. The kinematic data for visible stars would indicate strong increase of $M/L$ as in the case of M32. However X-ray luminosity from the accretion of tidally disrupted stars by black holes is likely to be very high (i.e., §2.3). Thus interpretation of steep rise in $M/L$ in M32 as a presence of black hole subcluster is highly unrealistic.

As we have seen from Figure 2, the nuclei with velocity dispersion smaller than $\sim 50\,km/sec$ cannot hold the three-body binaries, but they rather are ejected at some hardness. Since the three-body binary formation is very efficient in low velocity dispersion environment, this process is likely to be efficient in removing black holes, as discussed by Kulkarni, Hut & McMillan (1993) in the context of globular clusters. This may be the case for the central nucleus of M33, which has very little evidence for a massive black hole (Kormendy 1993).

## 5. Future Works and Conclusions

So far we have limited ourselves to simple models composed of two components: 0.7 $M_\odot$ main-sequence stars and 10 $M_\odot$ stars. The physical processes included were the dynamical relaxation, three-body binary heating, and the first generation mergers between the black holes. Each process has its own uncertainties, which may or may not affect the general evolution, as discussed below.

The three-body binary heating requires various kinds of close encounters. During complex three-body encounters, any two objects can become sufficiently close so that the gravitational radiation could seriously alter the outcome. This will affect both the efficiency of binary heating and the number of merger products. Thus more realistic modeling of three-body binaries tends to boost the production of massive black holes.

The recoil velocity during the black hole merger becomes very important (Redmount & Rees 1989) for the nuclei with escape velocity comparable to the recoil velocity itself. If the central potential is deep enough to hold the merger products, the black hole mass should grow continuously. However, if the merged black holes escape from the nucleus whose potential depth is shallow, the formation of a central black hole would be difficult, and the two-component structure of central subsystem composed of 10 $M_\odot$ black hole plus a surrounding system with low mass main-sequence stars may be long-lived. Unfortunately, the current estimates of recoil velocity appear to be very uncertain.

We have estimated that the tidal capture of the main-sequence stars by the black holes would provide a significant growth of black hole mass. The accretion of stellar material by the black holes should produce very strong radiation. This could be an argument against the possibility of having a cluster of black holes in nuclei of M32 or our galaxy, but better understanding of the star-black hole interaction is necessary.

We found that the condition for the buildup of mass can be easily fulfilled with much less extreme initial conditions than required for single component case. In order to model the runaway merger to a seed black hole, one has to perform the N-body calculation similar to that one performed by Lee (1993). Since the black hole subsystem is relatively small $N$ system and the background stellar system is nearly static, we may be able to model the dynamics with present computing power.

We have used the simple models for the initial conditions. Actual galactic nuclei should have range of masses for the main-sequence stars that will generate neutron stars, white dwarfs and black holes. It is true that the massive stars will quickly evolve off, but there are stars of intermediate mass that could stay for the period of substantial dynamical evolution. The evolutionary path outlined here should be altered somewhat depending on the shape of the mass function. We have assumed a single mass for the black hole but there should be a spread of black hole masses ranging from 5 to 20 $M_\odot$. The presence of spread in black hole will also have significant dynamical effects, but the main conclusion of the present study will not change much.

It is clear that the stellar dynamical process alone would not be able to form a seed black hole more massive than the core mass. The central core regulated by the heating by three body process will maintain the core composed of order of 100 black holes. This means that the seed black hole mass would not be much more massive than 1000 $M_\odot$. The question of the subsequent growth requires further knowledge on the environment. If there is enough amount of gas nearby, the black hole can grow fast by accreting the gas.

The possible existence of small amount of $10\,M_\odot$ black holes embedded in a cluster of lower mass stars would make the evolution of a spherical stellar system quite different from the cases without such a component. Most notable effect is the acceleration of the core collapse which would be halted only at a great density. The merger of three-body binaries and capture of black holes by gravitational radiation are likely to drive the formation of massive black holes during the oscillatory phase after the core bounce as a result of heating effect of three-body binaries. If there existed a stellar system with a velocity dispersion exceeding $\sim 500\,km/sec$ a seed black hole may even be formed during the first collapse.

I would like to thank G. Quinlan and M. H. Lee for critical comments, and D. Chernoff for useful conversations. This work was supported in part by the Korea Science and Engineering Foundation under grant No. 941-0200-001-2 and in part by the National Science Foundation under Grant No. PHY89-04035.